\documentstyle[12pt]{article}
\def\sst{\scriptscriptstyle}

\begin{document}
\begin{flushright}
{\small { MSUCL--884 }\\
\vskip -5pt
{ WSU--NP--93--1 }}
\end{flushright}
\bigskip
\thispagestyle{empty}
\begin{center}
{\large  {\bf NUCLEAR FLOW EXCITATION FUNCTION}}\\
  \bigskip
  \bigskip
{\large Dietrich Klakow,$^{a,*}$ Gerd Welke,$^b$ and Wolfgang Bauer$^a$}\\
\bigskip
{\small $^a$National Superconducting Cyclotron Laboratory\\
\vskip -5pt
and Department of Physics and Astronomy,\\
\vskip -5pt
Michigan State University, East Lansing, MI 48824.\\
\smallskip
\smallskip
$^b$Department of Physics and Astronomy,\\
\vskip -5pt
Wayne State University, Detroit, MI 48202.}\\
\bigskip
\end{center}
\bigskip
\bigskip
\begin{center}
\noindent {\large {\bf Abstract}}
\end{center}
We consider the dependence of collective flow on the nuclear surface thickness
in a Boltzmann--Uehling--Uhlenbeck transport model of heavy ion
collisions.  Well defined surfaces are introduced
by giving test particles a Gaussian density profile of constant width.
Zeros of the flow excitation
function are as much influenced by the surface thickness as
the nuclear equation of state, and the dependence of this effect is
understood in terms of a simple potential scattering model.
Realistic calculations must
also take into account medium effects for the nucleon--nucleon
cross section, and impact parameter averaging.
We find that balance energy scales with the mass number as $A^{-y}$,
where $y$ has a numerical value between 0.35 and 0.5, depending on the
assumptions about the in-medium nucleon-nucleon cross section.

\bigskip
\bigskip
\vfill
\noindent {\bf PACS Indices:} 25.70.-z, 02.70.Lq, 21.65.+f

\newpage

Broadly speaking, nuclear collective flow in a heavy ion collision
is the deflection of nuclear matter
perpendicular to the beam axis during the course of the reaction.
Experimentally, one observes that flow disappears at
a well defined beam energy~\cite{KRO}-\cite{WES2}, the so called balance
energy $E_{Bal}$, whose value depends on the system and impact
parameter range being considered.
These zeros in the flow excitation function were predicted
by the Boltzmann--Uehling--Uhlenbeck (BUU) transport
model~\cite{MOL,BER}, and an analysis of scale invariant
quantities~\cite{BON}, and may be understood as an overall cancellation of
the attractive part of the mean field interaction with repulsive
contributions from the mean field and collisional kinetic pressure.
Thus, as has been shown explicitly for
BUU simulations in Ref.~\cite{OGI}, $E_{Bal}$ is expected to
depend on both the nuclear equation of state and the magnitude of the
in--medium nucleon--nucleon cross section. By making a systematic
study of the balance energy as a function of the nuclear mass,
one therefore hopes to gain insight into these properties. However,
other parameters might well influence the balance energy. In particular,
we wish to investigate in this note the effect of finite nuclear
surface thicknesses and impact parameter variations
on $E_{Bal}$. We begin by defining the flow variable
to be used here, and point out the importance of obtaining well--defined
nuclear surfaces that are independent of the grid size used to compute
density gradients. We then show how the strong surface dependence of flow in a
Vlasov simulation may be understood in terms of a simple potential
scattering model. This dependence persists for full BUU calculations
that include a non--zero collision integral. Lastly, we consider the
effect on the balance energy when the mass number and impact parameter
are varied.

To analyze flow quantitatively in experiments one of the main
problems that has to be addressed is the determination of the reaction plane
(see, for example, Refs.~\cite{DAN1}--\cite{WIL}).
In a model calculations, on the other hand,
knowledge of the reaction plane immediately allows one to define a
flow variable such as the average in--plane transverse momentum
\begin{equation}\label{defflow}
\langle w P_x \rangle \; = \; {1 \over N}\, \sum_{i=1}^{N}\, w_i
P^x_i~ ~,
\end{equation}
\noindent where the weight $w_i=1$ or $-1$ if the test particle $i$
is emitted  into the
forward or backward center--of--mass hemispheres, respectively, and
$P^x_i$ is the transverse momentum of the test particle in the reaction plane.
In this note, we shall refer to ``flow'' in the sense of the above equation.

We begin by examining the way the local particle density $\rho$ is
calculated for the mean field dynamics in a Vlasov simulation.
While the mean field is momentum dependent~\cite{GALp}--\cite{GAL}, we
shall consider here, for illustrative purposes, a Skyrme-like
parametrization that is a function of the density alone:
\begin{equation}\label{U}
U(\rho)=A{\rho\over{\rho_0}}+B\left({\rho\over{\rho_0}}\right)^\sigma ~ ~,
\end{equation}
and take values of the parameters $A=-124~{\rm MeV}$, $B=70~{\rm MeV}$, and
$\sigma=2$. This choice reproduces known nuclear matter properties,
with a rather stiff compression modulus at saturation of $K_0=380~{\rm MeV}$.

We choose to represent the nucleon phase space distribution function
$f(\vec r,\vec p,t)$ by an ensemble of test particles. If
$f(\vec r,\vec p,t)$ is to satisfy the Vlasov equation,
the equations of motion of a test particle
with coordinates $(\vec r_i, \vec p_i)$ are given by Hamilton's
equations of motion with potential (\ref{U}):
\begin{equation}\label{eqn}
\dot{\vec p_i} =-{\vec \nabla_r} U(\rho({\vec r_i})) \qquad
{\rm and} \qquad
\dot{\vec r_i} = {{\vec p_i}\over \sqrt{m_{\sst N}^2+p_i^2} }~ ~,
\end{equation}
\noindent where $m_{\sst N}$ is the free nucleon mass.
The local particle density $\rho$ is often calculated on a grid, and
the gradient obtained as a finite difference.
In this procedure, each test particle
counts a certain fraction towards the density of the cell it occupies,
while neighboring cells receive a smaller contribution~\cite{BER2}.
This stabilizes the numerics, but also introduces a nuclear surface
whose thickness is roughly given by the grid size $\Delta x$.
This grid size must be larger than the maximum distance
traversed by a test particle in one time step, but small enough to be
able to compute the gradients in Eq.~(\ref{eqn}) to sufficient accuracy.

Instead of ``smearing'' a test particle in steps over only the
nearest neighboring cells, one may choose to supply each test particle
with, say, a Gaussian density profile of constant
width. This also introduces a finite surface thickness, but one that
is well--defined and independent of the grid size. Thus one may study
the effect of varying the nuclear surface thickness without the
external numerical constraints imposed on the choice of $\Delta x$.

As an illustrative example,
we consider flow in pure Vlasov dynamics as a function of the size
of the grid for $^{139}{\rm La}$ on $^{139}{\rm La}$ collisions at
a beam energy of 800~MeV/nucleon and impact parameter $b=2.7~{\rm fm}$.
Fig.~1 shows that $\langle w P_x \rangle$ depends
very strongly on the grid size, i.e., the nuclear surface (dotted line).
The extrapolated value to zero grid size (zero surface thickness)
is twice as large as the value at $\Delta x=1~{\rm fm}$, a frequently
used grid size. This strong dependence persists at beam energies of
200~MeV/nucleon. At this energy, calculations
with finite $\Delta x>0.5~{\rm fm}$ gave
an overall attraction, whereas the extrapolation to $\Delta x=0$
predicted positive flow.

In Fig.~1, we also show the flow obtained with Gaussian test
particle density profiles. The solid line represents the variation
with $\Delta x$ for a surface thickness of 2~fm, while the dashed line
is for a surface of 1~fm. Clearly, as long as the grid size
is smaller than the Gaussian smearing width (i.e. the
nuclear surface), the flow is independent of $\Delta x$.
Also, the dotted line crosses the solid and dashed lines at $\Delta x
\sim 2$ and 1~fm, respectively. This indicates that at least most of its rise
with decreasing grid size is directly attributable to the changing
surface thickness, and is not a ``numerical artifact.''
We conclude that $\langle w P_x\rangle$ depends rather strongly on the
surface thickness, a thinner surface
producing more than a thicker one.
Of course, quantitatively the results obtained here are not reliable, since we
have, for instance, ignored the momentum dependence and hard collisions,
but they do show the important influence of the nuclear surface.

The surface dependence of flow can be understood
in terms of a simple potential scattering model~\cite{GAL}.
Assuming that the nuclei  pass through each other without
changing their shape in phase space, the centers of mass of the nuclei
move according to the Hamilton function
\begin{equation}\label{H}
H={{P_1^2}\over{2M}}+{{P_2^2}\over{2M}}+\tilde{V}(R)~ ~,
\end{equation}
\noindent where the potential $\tilde{V}$ is given by
\begin{equation}\label{toypot}
\tilde{V}(R)=\int V(\rho_{\sst R}(r)) dr^3~ ~,
\end{equation}
\noindent $M$ is the nuclear mass, $P_1$ and $P_2$
the center--of--mass momenta, and $R$ the separation of the nuclei.
In (\ref{toypot}), $V(\rho_{\sst R})$ is the potential energy
density corresponding to Eq.~(\ref{U})
\begin{equation}\label{potden}
V(\rho) \; = \; {A\over2}{\rho^2\over\rho_0}+
{B\over{\sigma+1}}{\rho^{\sigma+1}\over\rho_0^\sigma}.
\end{equation}
\noindent To describe nuclei with a surface, we choose the density
profile to be
\begin{equation}\label{den}
{\rho_{\sst R}(r)\over\rho_0} \;=\;
 \bigg [{1+\exp{\left({|\vec{r}-\vec{R}/2|-R_0}\right)/a}} \bigg ]^{-1}
\:+\: \bigg [{1+\exp{\left({|\vec{r}+\vec{R}/2|-R_0}\right)/a}}
\bigg ]^{-1}~ ~,
\end{equation}
\noindent where $R_0$ is the nuclear radius, and $4a$ the surface thickness.

The potential (\ref{potden}) is shown in Fig. 2 for various values
of the parameter $a$.
As expected, for an increasing surface thickness, the potential
$\tilde{V}$ decreases, and we expect a larger
surface to produce less flow.
This can be shown explicitly in the time evolution
of the $\langle w P_x \rangle$ (see Fig. 3).
We find reasonable agreement of the potential scattering model with the
test particle Vlasov calculation at both values of the surface thickness.
Of course, differences are seen in the details, and are expected because of the
crude assumptions made in Eq.~(\ref{H}).
For example,
a Vlasov calculation shows that $E_{Bal}$ occurs between 200~MeV and 300~MeV,
depending on the surface, while the scattering model shows no zeros in
the flow excitation function.

We now focus on the disappearance of flow in a more realistic
calculation, i.e., a full BUU simulation that includes
collisions. For the reaction $^{139}{\rm La}$ on $^{139}{\rm La}$~\cite{KRO}
we find that the balance energy is shifted by $\approx 10~{\rm MeV}$
when the surface thickness increases from 1~fm to 2~fm.
This is comparable to the shift expected when one changes
from a stiff to a soft equation of state. For example,
in $^{40}{\rm Ar}$ on V reactions, $E_{Bal}$ changes by only $8~{\rm MeV}$
if the incompressibility is increased from $K_0=200~{\rm MeV}$ to
$380~{\rm MeV}$ by adjusting the parameters in Eq.~(\ref{U})~\cite{OGI}.

This becomes even more apparent in smaller systems such as $^{12}{\rm
C}+^{12}{\rm C}$, for which values of $E_{Bal}$ were recently
measured at the National Superconducting Cyclotron Laboratory~\cite{WES}.
Fig.~4 shows the flow obtained in simulations as a function of beam energy,
for surface thicknesses of 1~fm (left panel) and 2~fm (right panel).
The stiff EOS gives $E_{Bal}\approx140~{\rm MeV}$ and
$190~{\rm MeV}$, respectively, while the soft EOS yields
$E_{Bal}\approx170~{\rm MeV}$ and $220~{\rm MeV}$, respectively.~\cite{TRI}.
The relative importance of the surface thickness is enhanced in
smaller systems. Of course,
the use of momentum dependent mean fields, a local Thomas-Fermi
momentum space initialization, and values of the in--medium NN cross section
will also influence the balance energy.
For instance,
it has been shown~\cite{OGI} that BUU calculations with reduced
NN cross sections result in higher values of the balance energy.

Fig.~5 shows the balance energy as a function of the combined mass of
the system. Experimental data (squares and diamond) suggests that
the dependence of the balance energy on the combined mass, $A$,
of the colliding nuclei follows a power law,
\begin{equation}
\label{ebal}
E_{Bal} \; = \; x A^{-y}~ ~,
\end{equation}
where the exponent $y$ has a numerical value of $0.33\pm 0.04$.  This power law
dependence is reproduced by the BUU simulation
(triangles and circles correspond to a soft and stiff EOS
respectively)~\cite{note}. In this calculation,
we have parametrized the the nucleon--nucleon
cross sections
in terms of a least squares fit to the experimental data of Ref.~\cite{PDG}.
The resulting values are somewhat different from the iso--spin averaged
expressions used at higher beam energies.
In particular, $\sigma_{pp}\ne\sigma_{pn}$, and the
cross sections are larger
than the ones described in Ref.~\cite{BER2}. This results in a shift
of the balance energy to lower values.

In Fig.~6, we investigate the dependence of the balance energy on the
value of the in-medium nucleon-nucleon cross section, where we
look for medium corrections beyond the effect of the Pauli principle on
the outgoing scattering states.  In previous studies
the free nucleon-nucleon cross section was multiplied with an overall
constant scaling factor \cite{OGI,BER,MSF}.  However, this approach fails
when one has collisions in low-density nuclear matter, where the in-medium
cross section should approach its free-space value.  A more realistic approach
uses a Taylor expansion of the in-medium cross section in the density
variable:
\begin{eqnarray}
  \sigma_{\rm NN}(\sqrt{s},\rho) &=& \sigma_{\rm NN}(\sqrt{s},0) +
      \rho\left.{\partial\,\sigma_{NN}(\sqrt{s},\rho)\over\partial\rho}
          \right|_{\rho=0} + \ldots \nonumber\\
      &=& \left(1 + \alpha_1\,{\rho\over\rho_0}\right)
          \sigma_{\rm NN}(\sqrt{s},0) + \ldots
  \label{medium1}
\end{eqnarray}
where we have introduced the dimensionless parameter $\alpha_1$, given by:
\begin{equation}
  \label{medium2}
  \alpha_1 = \rho_0\,{\partial\over\partial\rho}
                     \left.\left\{\ln\sigma_{\rm NN}(\sqrt{s},\rho)
                         \right\}\right|_{\rho=0}\ .
\end{equation}
In principle, $\alpha_1$ is dependent on $\sqrt{s}$, but we have here -- as
a first approximation -- taken $\alpha_1$ as an energy-independent constant.
Fig.~6 shows the mass dependence of the calculated balance energy as a
function of different values of $\alpha_1$, where we have used a soft
nuclear equation of state.  It is clear that we obtain the
best overall agreement with the experimental data for a value of
$\alpha_1=-0.2$,
correponding to a 20\% reduction of the nucleon-nucleon cross section at
$\rho=\rho_0$.  However, the power-law exponent $y$ from Eq.\ (\ref{ebal})
depends on the choice of $\alpha_1$, with $\alpha_1 = 0$ (no medium
modification) yielding the best
agreement ($y=0.38\pm 0.05$) with the experimental value of $y=0.32$.
For $\alpha_1=-0.1$ one extracts $y=0.42\pm 0.03$, and for
$\alpha_1=-0.2$ we obtain $y=0.47\pm 0.03$.
That $y$ increases with $|\alpha_1|$ is due to the fact that at higher
beam energies higher densities are reached, and therfore the reduction
of the in-medium cross section is stronger for the lighter systems, which
have higher balance energies.  This effect may, however, be at least
partially compensated once one incorporates a more realistic energy dependence
of $\alpha_1$.

In addition to effects mentioned previously, the proportionality constant
$x$ (and, in principle, also the exponent $y$) in Eq.\ (\ref{ebal}) also
depends on the impact parameter $b$~\cite{TSA}, as can been seen in
Fig.~7. For peripheral collisions,
the contribution of the nucleon-nucleon collisions becomes relatively
less important for the flow production than the mean
field, because the repulsion generated by the nucleon-nucleon collisions
is proportional to the overlap volume.
We note that
the calculations in Figs.~5 and 6
use only one impact parameter, obtained from the average value in the
experiment, and proportionally scaled $b$
for calculations at mass numbers $A$ for which no data exists.
Of course, for a quantitative comparison with experiment a
impact parameter weighted averaged flow has to be considered, but
calculating $E_{Bal}$ for several $A$ and $b$ is computationally prohibitive.

In conclusion, we have shown that in Vlasov and BUU simulations
reliable results for flow must take into account the finite thickness
of the nuclear surface. This is best done
by giving the test particles
a Gaussian density profile, with a width
that is larger than the grid size used to obtain the
density gradients. The value of the nuclear
surface thickness has a strong effect on the balance energy. Therefore,
more quantitative BUU predictions have to
not only take into account the equation of state and medium
effects on the
nucleon-nucleon cross section,
but also proper initial conditions in phase space and impact parameter
averaging.
In addition, we find that a realistic variation of the in-medium
nucleon-nucleon cross-section with density has a clear effect on the mass
dependence of the balance energy.  We find best overall agreement with the
experimental data for $\alpha_1=-0.2$, where $\alpha_1$ is defined in
Eq.\ (\ref{medium2}).

\bigskip
\noindent {\bf Acknowledgements}

We are grateful to G.F. Bertsch and
B.-A. Li for fruitful discussions. This work was supported the
National Science Foundation
under Grant No. 90-17077.
W.B. acknowledges support from an NSF
Presidential Faculty Fellow award, and D.K. was partially supported by the
Studienstiftung des Deutschen Volkes.\\
\bigskip
\bigskip\\
$^*$Present address: Institut f\"{u}r theoretische Physik II,
Staudtstra{\ss}e 7, W-85200 Erlangen, Germany.

\newpage

\noindent {\Large {\bf Figure Captions}}

\bigskip

\noindent {\bf Fig.1}~ ~Flow (as defined in Eq.~(\ref{defflow})) versus
grid size for pure Vlasov dynamics
with smearing over neighboring cells only (dotted line).
The dashed and solid lines show results if each test particle is given
a Gaussian density profile of constant width.
The dashed line corresponds to a nuclear surface thickness
of 1~fm, the solid line to 2~fm. All curves shown are for
$^{139}{\rm La}+{\rm La}$, $E=800~{\rm MeV/nucleon}$, and $b=2.7~{\rm fm}$.

\bigskip

\noindent {\bf Fig.2}~ ~Potential
  Eq.~(\ref{toypot}) as a function of the separation of
  the two nuclei. The radius $R_0=5.8~{\rm fm}$ corresponds
  to $^{139}{\rm La}$.

\bigskip

\noindent {\bf Fig.3}~ ~The time
evolution of flow for pure
  Vlasov dynamics (solid line), and the potential scattering model,
  Eq.~(\ref{H}) (dashes).
  The surface thicknesses are 1~fm (upper two curves)
  and 2~fm (lower two curves), and the system is as for Fig.~1.

\bigskip

\noindent {\bf Fig.4}~ ~Flow versus beam energy in the vicinity of
  the zeros of the excitation function. The system is $^{12}{\rm
  C}+^{12}{\rm C}$ at an impact parameter
  $b=1.4~{\rm fm}$. The left panel is for
  a surface thickness of 1~fm, the right for 2~fm. Both panels
  show results for stiff (diamonds) and
  soft equations of state (plus signs).

\bigskip

\noindent {\bf Fig.5}~ ~The calculated
   values of $E_{Bal}$ as a function of the
   mass of the system. Only symmetric systems are considered.
   Diamonds and circles correspond to a soft and stiff EOS,
   respectively.
   For comparison, experimental data from Ref.~\cite{WES2}
   (squares) \cite{WES2}  are
   shown. The solid and dashed curves are power-law fits to the
   calculations and data.

\bigskip

\noindent {\bf Fig.6}~ ~Same as Fig.\ 5, but varying the in-medium cross
   section according to Eq.\ (\ref{medium1}) with $\alpha_1$ defined in
   Eq.\ (\ref{medium2}).  For all calculations, a soft equation of state
   was used.  The lines represent power-law fits to the calculations and
   data.

\bigskip

\noindent {\bf Fig.7}~ ~The balance
energy for ${\rm Cl}+{\rm Cl}$ as a function of impact parameter.
\end{document}